\documentclass[aps,prb,twocolumn,showpacs,amsfonts]{revtex4}

\usepackage{epsfig}
\usepackage{amsmath}
\usepackage{amssymb}
\usepackage{bm}

\newcommand{\beq}{\begin{equation}}
\newcommand{\eeq}{\end{equation}}
\newcommand{\beqarray}{\begin{eqnarray}}
\newcommand{\eeqarray}{\end{eqnarray}}

\newcommand{\eq}[1]{Eq.~(\ref{#1})} 
\newcommand{\fig}[1]{Fig.~\ref{#1}} 
\newcommand{\Ref}[1]{Ref.~\onlinecite{#1}} 

\begin{document}

\allowdisplaybreaks

\title{Excitonic spin density wave state in iron pnictides}
\author{P. M. R. Brydon}
\email{brydon@theory.phy.tu-dresden.de}
\author{C. Timm}
\email{carsten.timm@tu-dresden.de}
\affiliation{Institut f\"{u}r Theoretische Physik, Technische Universit\"{a}t
  Dresden, 01062 Dresden, Germany }

\date{\today}

\begin{abstract}
We examine the appearance of a spin density wave in the FeAs parent
compounds due to an excitonic instability. Using a realistic four-band model,
we show that the magnetic state depends very sensitively upon the details
of the band structure. We demonstrate that an orthorhombic distortion of the
crystal enhances the stability of the antiferromagnetic order.
\end{abstract}


\maketitle

{\it Introduction}. The superconductivity of materials containing
FeAs layers is currently receiving much attention. Like the
cuprates, these systems become superconducting upon chemical
doping of an antiferromagnetic (AF) parent compound, specifically
ReFeAsO (Re is a  rare earth ion) or AeFe$_{2}$As$_{2}$ (Ae is an
alkaline earth ion).~\cite{Kamihara2008,Rotter2008} Intriguingly,
the AF state occurs only in the presence of an orthorhombic
distortion of the crystal, which fixes the AF ordering
direction.~\cite{1111coupling,122coupling} The likely role of spin
fluctuations in producing the superconductivity has lead to
intense scrutiny of the AF phase. The relatively small value of
the moment at Fe
sites,~\cite{1111coupling,122coupling,McGuire2008} metallic
transport properties,~\cite{McGuire2008,Hu2008} and observations
of reconstructed Fermi surfaces,~\cite{magneto,ARPES} provide
strong evidence that the AF state is a spin density wave (SDW)
arising from the nesting of electron and hole Fermi
surfaces.~\cite{Mazin2008,Singh2008,Korshunov2008}

In analogy to Cr,~\cite{Rice1970} a theory of the SDW based upon
the excitonic pairing of electrons and holes has been
proposed.~\cite{Han2008,Chubukov2008,Vorontsov2008} It is
important to determine if this scenario is sufficient to explain
the AF state, or whether a more complicated multi-orbital approach
is
required.~\cite{Kuroki2008,Ran2009,Yu2008,Korshunov2008,Lorenzana2008}
As previous works have used a highly-idealized model of the
electronic structure,~\cite{Han2008,Chubukov2008,Vorontsov2008}
with only two Fermi surfaces instead of the likely four or
more,~\cite{Mazin2008,Singh2008} it is not clear if the excitonic
SDW can give the observed magnetic
ordering.~\cite{1111coupling,122coupling} Furthermore, the effect
of the orthorhombic distortion on such a state remains unknown. We
address these problems here by studying the appearance of the
excitonic SDW in a four band model of
LaFeAsO.~\cite{Korshunov2008} Using a mean-field theory, we show
that the SDW state is sensitively dependent upon the doping and
the details of the band structure.~\cite{Mazin2009} In particular,
we examine the response of the SDW phase to changes in the
ellipticity of the electron pockets, the relative size of the hole
pockets, and an orthorhombic distortion of the crystal.

\begin{figure}
\includegraphics[width=0.95\columnwidth,clip]{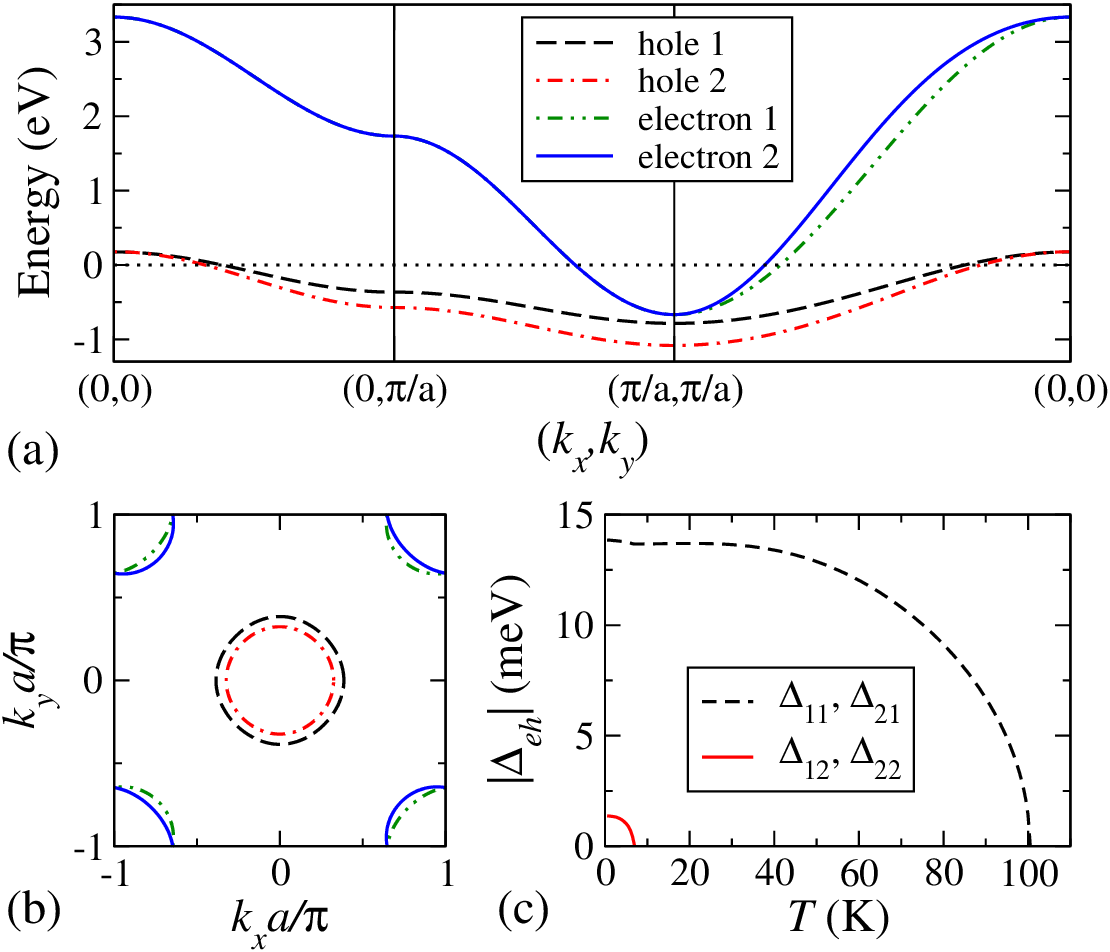}
\caption{\label{fig1}(color online) (a) Dispersion of the electron and hole
  bands along high-symmetry directions of the Brillouin zone. (b) Fermi
  surface at filling $n_{el}=4$. The bands are represented by the same curves
  as in (a). (c) Variation of the excitonic gaps with temperature at
  $n_{el}=4$.}
\end{figure}

{\it Theoretical model}. We model the FeAs planes as a 2D interacting
four-band system where two bands have electron-like Fermi surfaces and the
other two have hole-like Fermi surfaces. We write the Hamiltonian as
\beqarray
H &=&\sum_{n=1,2}\sum_{{\bf{k}},\sigma}\left\{\epsilon^{e}_{n{\bf{k}}}c^{\dagger}_{n{\bf{k}}\sigma}c^{}_{n{\bf{k}}\sigma}
+
\epsilon^{h}_{n{\bf{k}}}f^{\dagger}_{n{\bf{k}}\sigma}f^{}_{n{\bf{k}}\sigma}
\right\} \notag \\
& & +
\frac{1}{V}\sum_{n=1,2}\sum_{n'=1,2}\sum_{{\bf{k}},{\bf{k}}',{\bf{q}}}\sum_{\sigma,\sigma'}
\notag \\
&& \quad \times
\left\{g_{1}c^{\dagger}_{n,{\bf{k}}+{\bf{q}},\sigma}c^{}_{n{\bf{k}}\sigma}f^{\dagger}_{n',{\bf{k}}'-{\bf{q}},\sigma'}f^{}_{n'{\bf{k}}'\sigma'}
\right. \notag \\
&& \qquad +
g_{2}\left[c^{\dagger}_{n,{\bf{k}}+{\bf{q}},\sigma}c^{\dagger}_{n,{\bf{k}}'-{\bf{q}},\sigma'}f^{}_{n'{\bf{k}}'\sigma'}f^{}_{n'{\bf{k}}\sigma}
  \right. \notag \\
&&  \qquad\qquad +
  c^{\dagger}_{n,{\bf{k}}+{\bf{q}},\sigma}f^{\dagger}_{n',{\bf{k}}'-{\bf{q}},\sigma'}c^{}_{n{\bf{k}}'\sigma'}f^{}_{n'{\bf{k}}\sigma}
  \notag \\
&&  \qquad\qquad +
  f^{\dagger}_{n',{\bf{k}}+{\bf{q}},\sigma}c^{\dagger}_{n,{\bf{k}}'-{\bf{q}},\sigma'}c^{}_{n{\bf{k}}'\sigma'}f^{}_{n'{\bf{k}}\sigma}
  \notag \\
&& \left. \left.  \qquad\qquad +
  f^{\dagger}_{n',{\bf{k}}+{\bf{q}},\sigma}f^{\dagger}_{n',{\bf{k}}'-{\bf{q}},\sigma'}c^{}_{n{\bf{k}}'\sigma'}c^{}_{n{\bf{k}}\sigma}\right]\right\} \label{eq:Ham}
\eeqarray
where $c^{\dagger}_{n{\bf{k}}\sigma}$ ($f^{\dagger}_{n{\bf{k}}\sigma}$)
creates a spin-$\sigma$ electron with momentum ${\bf{k}}$ in the electron-like
(hole-like) band $n$.
Due to the out-of-plane arrangement of the As ions, the crystallographic
unit cell of the FeAs plane contains two Fe ions.  Our band structure
is given in terms of this unit cell, but in the
discussion of magnetic properties it is more useful to refer only to the Fe
lattice, which requires us to ``unfold'' the Brillouin zone.~\cite{Mazin2008}
Assuming crystallographic unit-cell
dimensions $a\times{a}$, the bands with electron-like Fermi
surface have dispersion
$\epsilon^{e}_{n{\bf{k}}} = \epsilon_{e} +t_{e,1}[\cos(k_{x}a) + \cos(k_{y}a)] +
t_{e,2}\cos([k_{x} + (-1)^{n}k_{y}]a/2)$, while for the hole-like bands we have
$\epsilon^{h}_{n{\bf{k}}} = \epsilon_{h,n} + t_{h,n,1}[\cos(k_{x}a) +
  \cos(k_{y}a)] + t_{h,n,2}\cos(k_{x}a)\cos(k_{y}a)$. In units of eV, we use
$\epsilon_{e}=1.544$, $t_{e,1}=1.0$, $t_{e,2}=-0.2$,
$\epsilon_{h,1}=-0.335$, $t_{h,1,1}=0.24$, $t_{h,1,2}=0.03$,
$\epsilon_{h,2}=-0.512$, $t_{h,2,1}=0.315$, and $t_{h,2,2}=0.06$.
We keep only the bands which intersect the Fermi surface. For
electron filling $n_{\mathrm{el}}=4$, corresponding to the undoped
parent compounds, we find the dispersion and Fermi surface as
shown in Fig.s~\ref{fig1}(a) and (b), respectively. Note that the
nesting of the hole and electron Fermi surfaces is not perfect,
since both the shape and the enclosed area differ. Our model
reproduces the Fermi surface and low-energy velocities of the band
structure proposed in~\Ref{Korshunov2008} for LaFeAsO, but
unlike~\Ref{Korshunov2008} obeys the correct periodicity of the
Brillouin zone.

The interaction terms in~\eq{eq:Ham} describe a density-density interaction
and correlated transitions between the electron and
hole bands, with contact potentials $g_{1}>0$ and $g_{2}>0$,
respectively. At low temperatures, the system is unstable against the pairing
of electrons and holes, producing an excitonic
state.~\cite{Rice1970,Excitonic,Kopaev1970}
Although
a rich variety of excitonic phases are possible, here the SDW
state has the largest effective coupling constant
$g_{s}=g_{1}+2g_{2}$.~\cite{Chubukov2008,Buker1981} At mean-field level, we therefore
decouple the interaction terms via the introduction of the real SDW excitonic
averages $\Delta_{eh} = (g_s/V)\sum_{\bf{k}}\sum_{\sigma}\sigma\langle{c^{\dagger}_{e,{\bf{k}}+{\bf{Q}},\sigma}f^{}_{h{\bf{k}}\sigma}}\rangle$
where ${\bf{Q}}=(\pi/a,\pi/a)$ is the nesting vector [see~\fig{fig1}(b)] and
$e (h)$ takes values of $1$ or $2$ to index the
electron (hole)
bands. $\Delta_{eh}$ is regarded as the order parameter of the
SDW state,~\cite{Rice1970,Excitonic} although it is only indirectly related to
the staggered magnetization.~\cite{Buker1981}
As each electron pocket is mapped to a different X point of the
enlarged Brillouin zone upon unfolding,~\cite{Mazin2008} the
$\Delta_{1h}$ and $\Delta_{2h}$ involve 
orthogonal nesting vectors ${\bf{Q}}_{1}$ and ${\bf{Q}}_{2}$ with respect to
the Fe sites, respectively. When both
$\Delta_{1h}$ and $\Delta_{2h}$ are non-zero, therefore, the magnetization is
the superposition of two orthogonal SDW states, each with stripe-like
ordering.~\cite{Lorenzana2008}

After decoupling the interaction terms, we obtain the equilibrium
mean-field solution by numerical minimization of the free energy
$F$ with respect to the $\Delta_{e,h}$. This was calculated over
the 2D Brillouin zone with at least a $1000\times1000$
${\bf{k}}$-point mesh. Throughout this work we set the effective
SDW coupling constant to be $g_{s}=0.84$eV, as at $n_{el}=4$ this
gives a partially-gapped Fermi surface in the SDW state with
reasonable critical temperatures: as shown in~\fig{fig1}(c) we
find that $\Delta_{e1}$ is non-zero below
$T_{\mathrm{SDW1}}=100$K, while $\Delta_{e2}$ appears below
$T_{\mathrm{SDW2}}=6.5$K. When all four averages are non-zero, we
find the inequality
$\Delta_{11}\Delta_{12}\Delta_{21}\Delta_{22}<0$; when only two
$\Delta_{eh}$ are present, their signs are independent.

\begin{figure}
\includegraphics[width=0.95\columnwidth,clip]{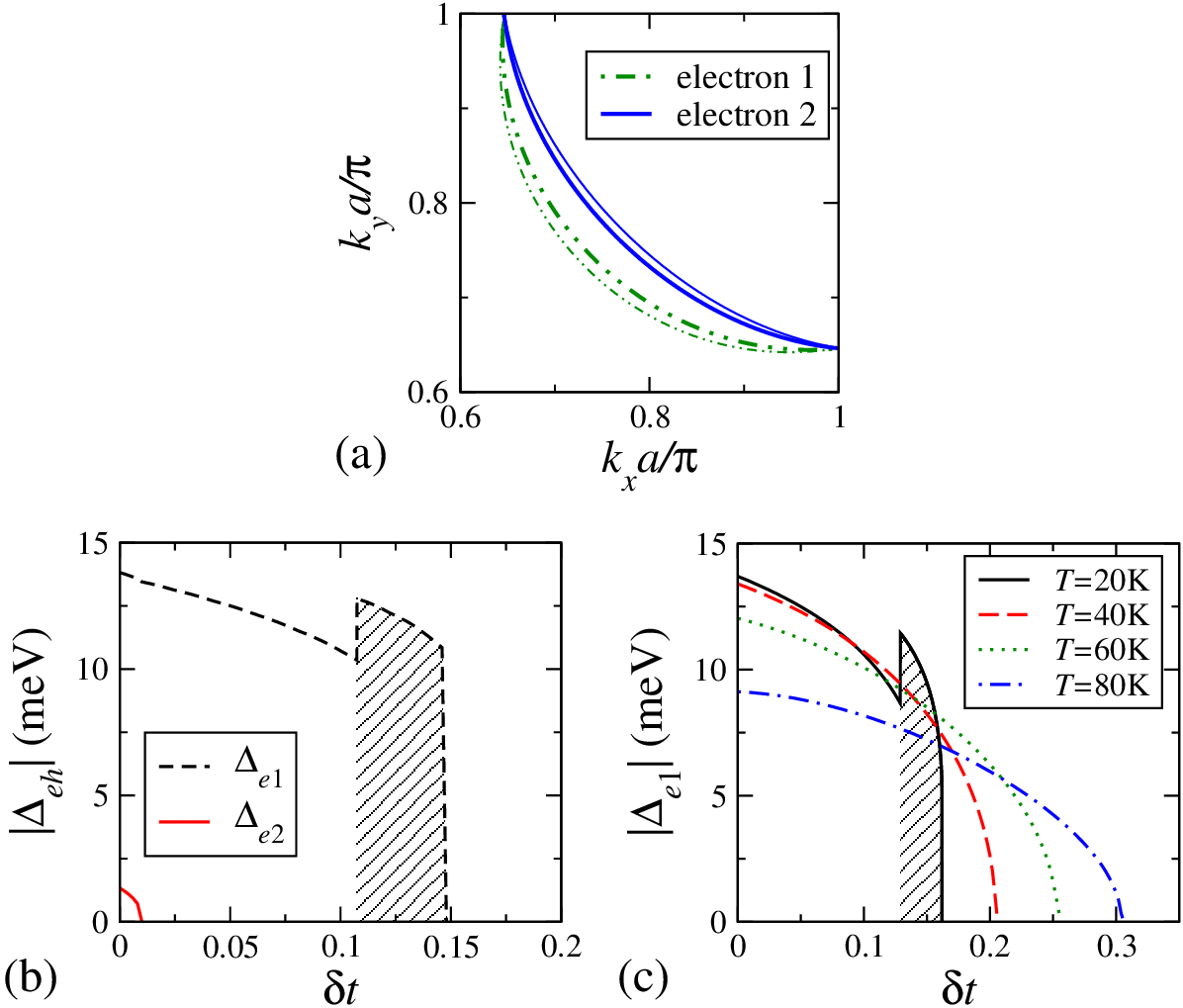}
\caption{\label{fig2}(color online) (a) Electron Fermi surfaces at
  $\delta{t}=0.2$ (thick lines) compared to $\delta{t}=0$ (thin
  lines). (b) Variation of $\Delta_{e,h}$ with $\delta{t}$ at $T=1$K. (c)
  Variation of $\Delta_{e,1}$ with $\delta{t}$ at various
  temperatures. Shading beneath the curve indicates a single-${\bf{Q}}$
  solution show in (b) and (c).}
\end{figure}

{\it Ellipticity of the electron pockets}. As seen in~\fig{fig1}(c), both
electron bands
participate in the excitonic instability at $n_{\mathrm{el}}=4$. This corresponds to
a ${\bf{Q}}_{1}+{\bf{Q}}_{2}$ SDW, whereas only a single-${\bf{Q}}$
SDW is experimentally observed.~\cite{1111coupling,122coupling} It has
previously been noted that these two
SDW phases should lie at similar energies,~\cite{Yu2008} and so it is
interesting to see
whether slight changes in the band structure can stabilize a
single-${\bf{Q}}$ state. This might be achieved, for example, by reducing the
ellipticity of the two electron pockets so as to enhance their competition for
the same states in each hole band. We therefore modify the electron dispersions
$\epsilon^{e}_{n{\bf{k}}} \rightarrow \epsilon^{e}_{n{\bf{k}}} +
2(-1)^{n}t_{e,2}\delta{t}\sin(k_{x}a/2)\sin(k_{y}a/2)$,
where the dimensionless parameter $\delta{t}$ controls the
ellipticity of the electron pockets. We compare the electron
pockets at $\delta{t}=0.2$ and $\delta{t}=0$ in~\fig{fig2}(a).

We find that even very small values of $\delta{t}\neq0$ can qualitatively
alter the
mean-field state. The evolution of the $\Delta_{eh}$ with
increasing $\delta{t}$ at $T=1$K is plotted in~\fig{fig2}(b). Reducing the
ellipticity of the electron
pockets tends to suppress the excitonic state, with $\Delta_{e2}$
disappearing before $\delta{t}=0.01$ is reached.
At $\delta{t}=0.108$ the system undergoes a
first-order transition from the ${\bf{Q}}_{1}+{\bf{Q}}_{2}$ state into a
single-${\bf{Q}}$ state. A single-${\bf{Q}}$ state is hence possible at
mean-field level by subtle modification of the band structure. Note
that the single-${\bf{Q}}$ states with
nesting vector ${\bf{Q}}_{1}$ and ${\bf Q}_2$ are degenerate.~\cite{Mazin2009}
Further increasing $\delta{t}$, the system undergoes a first-order transition
into the nonmagnetic state at $\delta{t}\approx0.145$.

The variation of $\Delta_{e,1}$ with $\delta{t}$ at higher temperature is
shown in~\fig{fig2}(c); in all cases $\Delta_{e,2}=0$. The first-order
transition
from the ${\bf{Q}}_{1}+{\bf{Q}}_{2}$ into the single-${\bf{Q}}$ state only
survives up to $T\sim30$K; at higher temperatures, the nonmagnetic state is
reached
from the ${\bf{Q}}_{1}+{\bf{Q}}_{2}$ phase by a second-order
transition. Interestingly, we see that the critical value of
$\delta{t}$ increases with $T$, even as the value of $\Delta_{e1}$ at
$\delta{t}=0$ is suppressed. This re-entrant behaviour 
is a generic feature of the phase diagram of the excitonic
insulator,~\cite{Rice1970,Kopaev1970} and may
indicate the presence of a low-$T$ incommensurate SDW
state.~\cite{Rice1970,Vorontsov2008}

\begin{figure}
\includegraphics[width=0.95\columnwidth,clip]{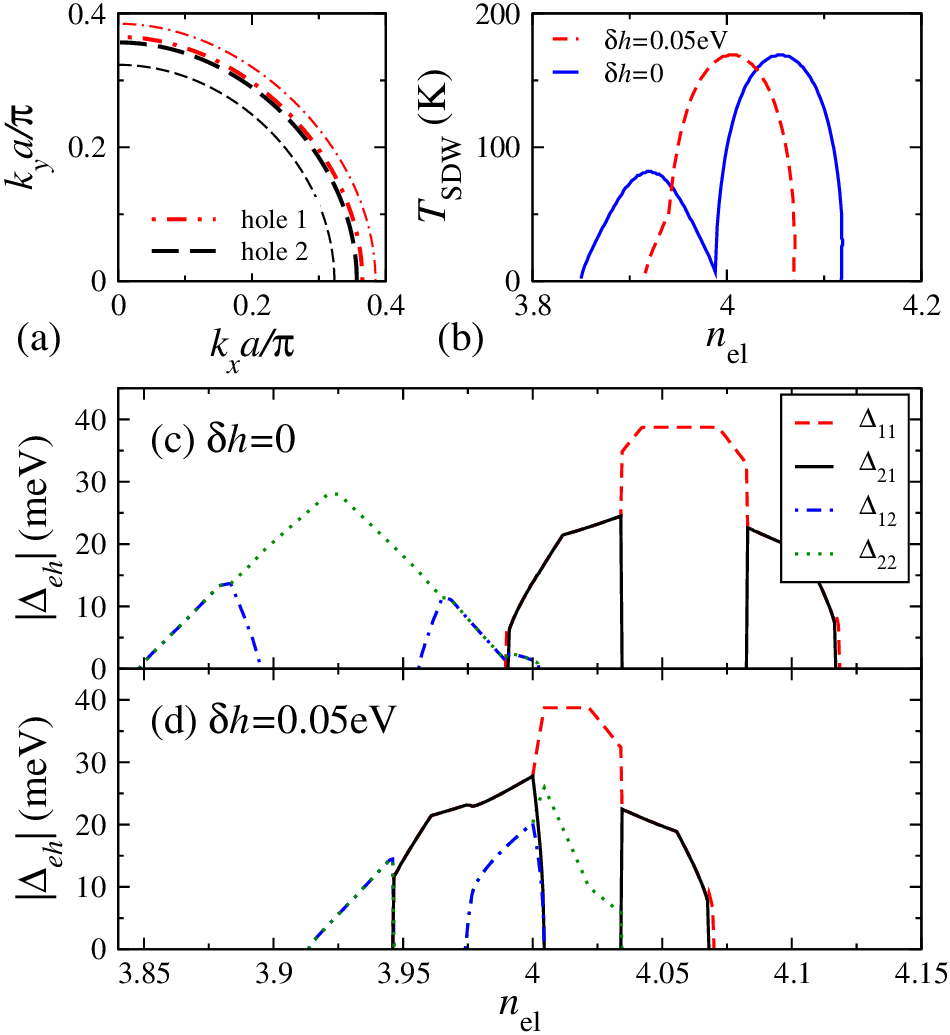}
\caption{\label{fig3}(color online) (a) Hole Fermi surfaces at
  $\delta{h}=0.05$eV (thick lines) compared to $\delta{h}=0$ (thin
  lines) at $n_{\mathrm{el}}=4$. (b) Dependence of the critical temperature
  $T_{\mathrm{SDW}}$ of
  the excitonic phases on electron filling $n_{\mathrm{el}}$ for different
  values of
  $\delta{h}$. (c) Variation of the $\Delta_{eh}$ with $n_{\mathrm{el}}$ at
  $T=1$K and
  $\delta{h}=0$. (d) Same as (c) but for $\delta{h}=0.05$eV.}
\end{figure}

{\it Hole pocket disparity}. The SDW state is sensitively
dependent not only upon the shape, but also upon the size of the
Fermi surfaces. This can be demonstrated in two ways: by raising
the energy of the second hole band
$\epsilon^{h}_{2,{\bf{k}}}\rightarrow\epsilon^{h}_{2,{\bf{k}}} +
\delta{h}$ so that the two hole Fermi surfaces converge  together,
or by varying the filling $n_{\mathrm{el}}$ to improve the nesting
between one of the hole Fermi surfaces and the two electron
pockets. At $\delta{h}=0.05$eV, the two hole Fermi surfaces are
nearly coincident when $n_{\mathrm{el}}=4$, see~\fig{fig3}(a). As
shown in~\fig{fig3}(b), this energy shift strongly alters the
$n_{\mathrm{el}}$-dependence of the maximum temperature
$T_{\mathrm{SDW}}$ at which at least one $\Delta_{eh}$ non-zero.
When $\delta{h}=0$, our model displays two distinct peaks in the
$T_{\mathrm{SDW}}$ vs.~$n_{\mathrm{el}}$ curve, with a sharp
minimum at $n_{\mathrm{el}}\approx3.99$. This behaviour
qualitatively disagrees with experiment, which shows only
monotonic suppression of $T_{\mathrm{SDW}}$ with
electron-doping.~\cite{Kamihara2008} The behaviour of
$T_{\mathrm{SDW}}$ at $\delta{h}=0.05$eV is in much better
agreement with experiment, with only a single maximum. Note that
the maximum value of $T_{\mathrm{SDW}}$ in both cases is
comparable to that in the ReFeAsO systems.~\cite{1111coupling}

The $T_{\mathrm{SDW}}$ vs.~$n_{\mathrm{el}}$ curves can be
understood by examining the evolution of the $\Delta_{eh}$ with
$n_{\mathrm{el}}$ at $T=1$K, plotted in~\fig{fig3}(c) for
$\delta{h}=0$ and in~\fig{fig3}(d) for $\delta{h}=0.05$eV. Note
that the values of the pairs $(\Delta_{11},\Delta_{22})$ and
$(\Delta_{21},\Delta_{12})$ may be swapped at every point. At
$\delta{h}=0$, the two distinct peaks in~\fig{fig3}(b) correspond
to a maximum in $|\Delta_{e1}|$  for electron doping and in
$|\Delta_{e2}|$ for hole doping. The maximum values are different
due to different densities of states in the hole bands. These
maxima occur when the area enclosed by the hole Fermi surface is
the same as that enclosed by each electron Fermi surface. It is
interesting to note that at both maxima a single-${\bf{Q}}$ state
is stable.

When $\delta{h}=0.05$eV, the conditions for $|\Delta_{e1}|$ and
$|\Delta_{e2}|$ to display a maxima coincide at $n_{\mathrm{el}}=4$, as the area
enclosed by each hole Fermi surface is almost equal. We hence see a
complicated coexistence between the four order parameters: at weak hole
doping, all four $\Delta_{eh}$ are non-zero; at weak electron doping,
the excitonic instability of the two electron bands involve different hole
bands.
Although $\Delta_{e1}$ is dominant over most of the doping range,
at extreme hole doping a state with only $\Delta_{e2}$ non-zero is realized,
corresponding to
the weak asymmetry seen in the $T_{\mathrm{SDW}}$ vs.~$n_{\mathrm{el}}$ curve
in~\fig{fig3}(b).

{\it Orthorhombic distortion}. In all known FeAs parent compounds,
the SDW phase occurs only in the presence of an orthorhombic
distortion of the crystal. It is found that the stripe-like SDW
has its nesting vector ${\bf{Q}}$ oriented along the longer
crystal axis.~\cite{1111coupling,122coupling} Here we see how this
can be understood within our model on the basis of the effect of
the orthorhombic distortion on the Fermi surfaces.

\begin{figure}
\includegraphics[width=0.95\columnwidth,clip]{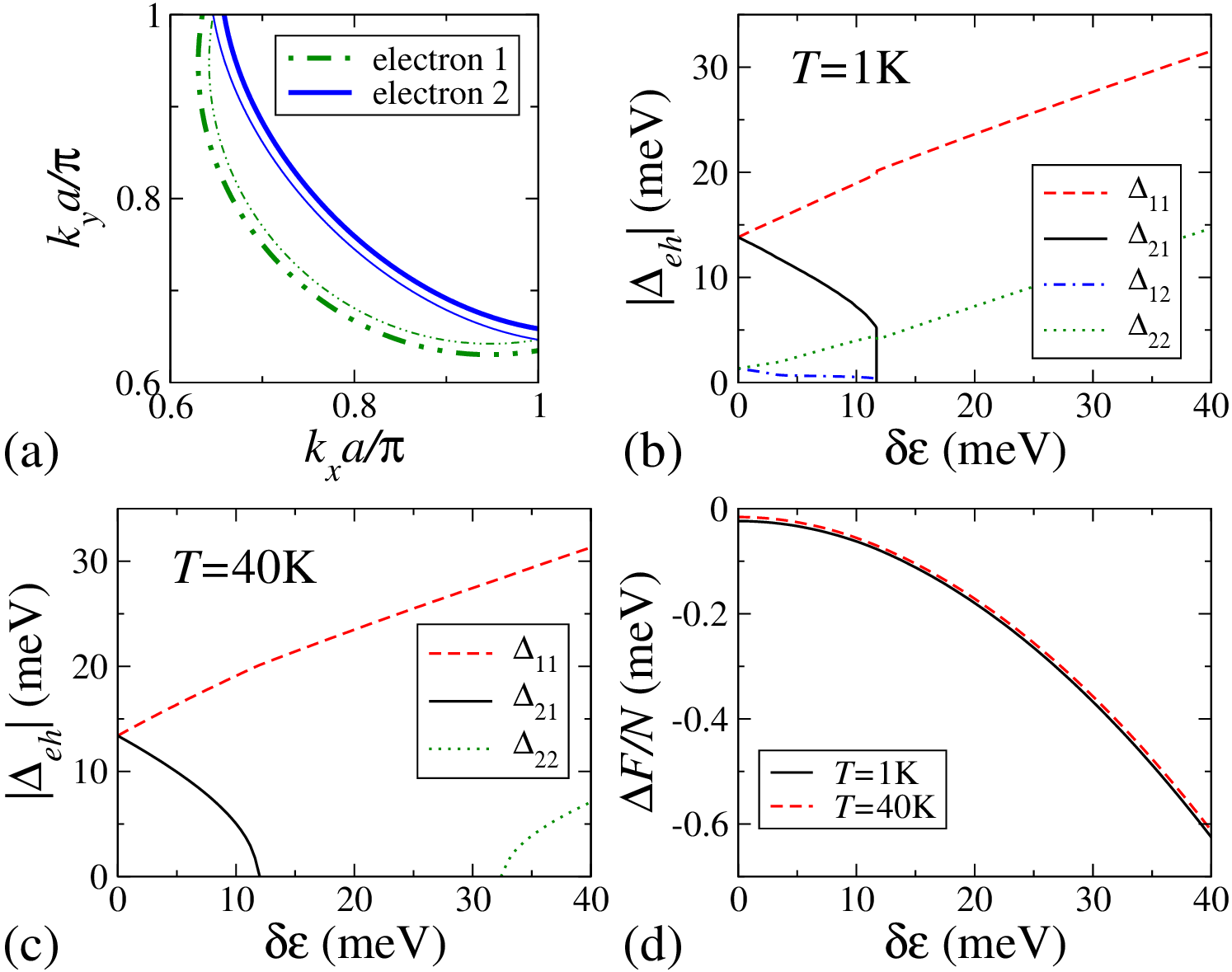}
\caption{\label{fig4}(color online) (a) Electron Fermi surface pockets at
  $\delta\epsilon=0.04$eV (thick lines) compared to $\delta\epsilon=0$ (thin
  lines). (b) Dependence of the $\Delta_{eh}$ upon $\delta\epsilon$ at
  $T=1$K. (c) Same as for (b) but at $T=40$K. (d) Difference
  $\Delta{F}=F-F_{0}$ per site between the
  free energy $F$ and its value in the normal state
  $F_{0}$ at $\delta\epsilon=0$.}
\end{figure}

Under an orthorhombic distortion, the energy shift of a state with
wave-vector ${\bf{K}}$ in the unfolded Brillouin zone is
$\delta\epsilon_{\bf{K}}\sim{\sum_{\alpha,\beta}K_{\alpha}Y_{\alpha,\beta}K_{\beta}}$
where $Y_{\alpha,\beta}$ is the strain tensor and we have
$Y_{xx}=-Y_{yy}$ and $Y_{xy}=0$.~\cite{Ziman} Note that the
wave-vectors ${\bf K}$ in the unfolded Brillouin zone are rotated
by 45$^{\circ}$ with respect to the wave-vectors in the
crystallographic Brillouin zone. We approximate the energy shifts
$\delta\epsilon_{\bf{K}}$ by their value near the chemical
potential, as the Fermi surface shape dominates the physics of our
model. The energy shifts of the hole states near the zone centre
are therefore neglected, as they will be much smaller than those
experienced by the electron pockets. Since the electron pockets
are small, we assume that their energy shifts are isotropic.
Furthermore, the sign of the energy shift will be opposite for the
electron pockets at the X points along the axes of compression
(negative energy shift) and dilation (positive energy
shift).~\cite{Ziman} We hence model the effect of the orthorhombic
distortion by
$\epsilon^{e}_{n{\bf{k}}}\rightarrow\epsilon^{e}_{n{\bf{k}}} +
(-1)^{n}\delta\epsilon$. We compare the electron pockets at
$\delta\epsilon=0.04$eV and $\delta\epsilon=0$ in~\fig{fig4}(a).

The dependence of the $\Delta_{eh}$ upon $\delta\epsilon$ at
$T=1$K and $T=40$K is plotted in~\fig{fig4}(b) and~\fig{fig4}(c),
respectively. The effect
of $\delta\epsilon\neq0$ is to enhance the pairing between the larger
electron
and hole pockets ($\Delta_{11}$) and also the smaller electron and hole
pockets ($\Delta_{22}$), while suppressing the pairing between the smaller
electron (hole) and larger hole (electron) pockets.
In analogy to the effect of doping, this can be readily understood as due to
the changes in the area enclosed by each electron Fermi surface.
Due to the enhanced excitonic pairing, the free energy $F$
shows monotonic decrease with increasing $\delta\epsilon$, see~\fig{fig4}(d). As
the orthorhombic distortion should increase the elastic energy of the lattice,
it is therefore possible that the total free energy of the crystal
will show a minimum at a non-zero value of the distortion. Deeper
investigation of this scenario is left for future work.

The $T=40$K case shows a large range of $\delta{\epsilon}$ where
$\Delta_{11}$ is the only non-zero excitonic average, i.e. the distortion
stabilizes a single-${\bf{Q}}$ SDW state due to the enhanced nesting
between the larger electron and hole pockets. In contradiction to experiment,
however, the ${\bf{Q}}$ vector is oriented along the shorter crystal
axis. This does not necessarily invalidate the excitonic scenario: our
model~\eq{eq:Ham} has equal coupling constants between the different bands. Were hole
band 2 to interact more strongly with the electron bands than hole band 1, so
that $|\Delta_{e2}|\gg|\Delta_{e1}|$ in the undistorted system, the
enhancement (suppression) of $\Delta_{22}$ ($\Delta_{12}$) by the orthorhombic
distortion would likely
stabilize a SDW state with the observed ${\bf{Q}}$ vector.

{\it Conclusions}. We have presented a mean-field study of the
excitonic SDW state for a realistic four-band model of the FeAs
parent compounds. We find that the SDW state is sensitively
dependent upon the band structure. For a tetragonal unit cell, a
two-${\bf Q}$ SDW is realized at $n_{\mathrm{el}}=4$; small
changes in the electron pocket ellipticity or the doping, however,
stabilize the observed single-${\bf Q}$ state. Varying the
relative size of the hole pockets qualitatively changes the
$T_{\mathrm{SDW}}$ vs.~$n_{\mathrm{el}}$ curve, agreeing best with
experiment when the hole pockets are almost
coincident.~\cite{Kamihara2008,1111coupling} The dominant effect
of an orthorhombic distortion of the crystal on the band structure
was identified as altering the size of the electron pockets. This
changes the nesting condition between the Fermi surfaces, and can
realize a single-${\bf Q}$ SDW. Our analysis suggests that the
electron pockets interact more strongly with the smaller hole
Fermi surface than with the larger. We conclude that the excitonic
SDW model is capable of qualitatively describing the AF phase of
the FeAs parent compounds. The strong sensitivity of the SDW state
upon the band structure, however, shows that a quantitative
description requires a more detailed understanding of the
electronic structure than is currently available.

The authors thank I. Eremin and D. V. Efremov for useful discussions.

\end{document}